\documentclass{article}
\usepackage{arxiv}

\usepackage{balance}       
\usepackage{graphicx}     
\usepackage{graphics}      
\usepackage[T1]{fontenc}   
\usepackage{url}
\usepackage{color}
\usepackage{booktabs}
\usepackage{textcomp}
\usepackage{enumitem}
\usepackage{subfig}

\usepackage{ccicons}          

\title{Wikipedia and Westminster: Quality and Dynamics of Wikipedia Pages about UK Politicians}

\author{
	Pushkal Agarwal\\
	King's College London\\
	London, UK\\
	\texttt{pushkal.agarwal@kcl.ac.uk} \\
	\And
	Miriam	Redi\\
	King's College London\\
	London, UK\\
	\texttt{miriam.redi@gmail.com}\\
	\And
	Nishanth Sastry\\
	King's College London\\
	London, UK\\
	\texttt{nishanth.sastry@kcl.ac.uk}\\
	\And
	Edward Wood\\
  	UK Parliament\\
 	London, UK\\
	\texttt{woode@parliament.uk}\\
	\And
	Andrew Blick\\
	King's College London\\
	London, UK\\
	\texttt{andrew.blick@kcl.ac.uk}
}
\begin{document}
\maketitle

\begin{abstract}
Wikipedia is a major source of information providing a large variety of content online, trusted by readers from around the world. Readers go to Wikipedia to get reliable information about different subjects, one of the most popular being living people, and especially politicians. While a lot is  known about the general usage and information consumption on Wikipedia, less is known about the life-cycle and quality of Wikipedia articles in the context of politics. The aim of this study is to quantify and qualify content production and consumption for articles about politicians, with a specific focus on UK Members of Parliament (MPs).
 
First, we analyze spatio-temporal patterns of readers' and editors' engagement with MPs' Wikipedia pages, finding huge peaks of attention during election times, related to signs of engagement on other social media (e.g. Twitter). Second, we quantify editors' polarisation and find that most editors specialize in a specific party and choose specific news outlets as references. Finally we observe that the average citation quality is pretty high, with  statements on `Early life and career' missing citations most often (18\%). 
\end{abstract}



\maketitle

\section{Introduction}



Political communication is defined by Brian McNair as ``purposeful communication about politics''~\cite{mcnair2017introduction}. It includes (inter alia) communication about politicians and other political actors, and their activities, as contained in news reports, editorials, and other forms of media discussion of politics such as large-scale online encyclopedias like Wikipedia.  
Information about politics on Wikipedia would appear to fulfil at least three of the functions of the communication media in `ideal-type' democratic societies as described by McNair: informing citizens of what is happening around them; educating as to the meaning and significance of the `facts'; and publicising the activities of governmental and political institutions~\cite{mcnair2017introduction}. 

Wikipedia's political content is pretty extensive. For example, all 650 elected  Members of the Parliament (MPs) in the United Kingdom (UK) 
have a Wikipedia page or article\footnote{We use the terms `page' and `article' interchangeably. The dataset we collected from these articles is available for the research community here: ~\url{tiny.cc/wikipedia-mps}}. Such content is also widely reused by the broader web: due to the popularity of Wikipedia as a platform, the top results returned by search engines queried for MPs show links to their Wikipedia pages. 

Given the visibility of Wikipedia, and the importance of the online encyclopedia in forming public opinion, the integrity and completeness of its content is crucial, especially 
during ``times of shock''such as elections or referenda ~\cite{zhang2019participation,10.1145/1240624.1240698}. To ensure information quality, Wikipedia editors operate in compliance with the core content policies of Neutrality and Verifiability\footnote{\url{https://en.wikipedia.org/wiki/Wikipedia:Core_content_policies}}. However, the scale of information and the free-to-edit Wikipedia policy sometimes limit the capability of communities to maintain the quality and neutrality of Wikipedia pages, and little is known about the life-cycle and quality of Wikipedia articles in the context of politics.

In this paper, we contribute to political communication studies by investigating, for the first time, citizens' engagement with politicians through the lens of the largest online encyclopedia. 
We perform a large-scale quantitative analysis of collective attention, quality, and polarization on Wikipedia pages of Members of Parliament in the UK. To this end, we collect a huge dataset of all edits (231k) and views (160M, since 2015) to UK MPs' Wikipedia articles.
We first investigate the spatio-temporal patterns of readers' and editors' engagement using the 650 UK MPs' Wikipedia articles. To understand the value of the political information viewed by millions of readers every month, and identify potential biases in political content in Wikipedia, we next focus on the \textit{polarisation} of MP pages' editors and citations
. To do so, we look at editors' preferences in terms of the political party of the subjects they edit, as well as the overall polarisation of the sources across MP Wikipedia articles, and 
finally compute at the quality and completeness of the citations in MP Wikipedia articles. 
We list these focus areas in the following research questions:

\noindent\textit{\textbf{RQ-1} What are the spatio-temporal patterns on MPs Wikipedia pages?}  Are there times of heavy edit and pageview loads?
(\S\ref{sec:dynamics})

\noindent\textit{\textbf{RQ-2} What is the quality of edits and citations on these pages? }
Is there polarisation of edits and citations along party lines? What is the citation quality?
(\S\ref{sec:polarization})



We find that engagement with MP Wikipedia articles is synchronized with election periods, and that  a mild form of polarisation exists in editors' preferences and article references. Furthermore, we verify that citation quality is generally high.

Collectively, these findings contribute to an in-depth understanding of political communication via Wikipedia. We discuss potential implications for information quality monitoring and disinformation studies in \S\ref{sec:discussion}.
 


\section{Background and Related Work}

We now provide some brief background on the UK political system and then discuss related research work on different themes.

\textbf{The UK House of Commons.} The UK Parliament has two chambers, of which the House of Commons is the lower chamber. It comprises 650 Members of Parliament (MPs) who are elected to represent constituencies by the First-Past-The-Post (FPTP) system. Due to the operation of FPTP, most MPs belong to one of two main parties, Conservative and Labour.
The UK Parliament created its first website in 1996, featuring a range of parliamentary papers, including Hansard~\cite{hansard}, the near-verbatim report of proceedings of both houses. 
The parliament website now has extensive information about individual MPs and how parliament works~\cite{ukmembers}.


\textbf{Editors' Engagement with Wikipedia}
Wikipedia is a platform where thousands of volunteers revise and add content constantly~\cite{iba2010analyzing}.  The work which is most closely relevant to ours is a recent study which looked at Wikipedia pages of members of Germany's parliament (Bundestag)~\cite{gobel2018political, reif2017politicians}. They show that German MPs use Wikipedia pages to enhance their image. Changes in pages are made at regular intervals, and significant peaks in the number of edits are associated with pursuing re-election. In ~\cite{gobel2018political}, authors use IP based public user edits to identify edits made from within the Bundestag and characterise types of changes. 



\textbf{Wikipedia pages quality.}
There are several works on various dimensions of page quality. Certain editors proclaim their political leaning and form communities~\cite{neff2013jointly}. Polarised teams--those consisting of a balanced set of politically diverse editors--may create articles of higher quality than politically homogeneous teams~\cite{shi2019wisdom}. 
Citations play a major role in the monitoring of information quality on Wikipedia. To help with this, the \emph{Citation Need} classifier by ~\cite{Redi:2019:CNT:3308558.3313618} determines if a statement requires a citation.


To the best of our knowledge, this is the first study looking at political communication in Wikipedia in the context of UK Members of Parliament. 
While ~\cite{gobel2018political} focuses on edits coming from inside the German Parliament, our study focuses on a much larger group of edits, and for the first time studies partisanship and neutrality of the Wikipedia editors of the MP wiki pages. 
\section{Dataset}\label{sec:dataset}
\begin{figure*}[htb]
    \centering
    \subfloat[Dataset Counts]{
      \includegraphics[width=0.3\textwidth]{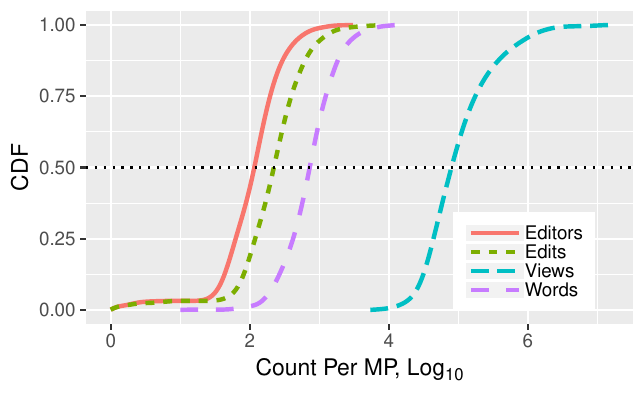}
      \label{fig:stats}
    }
    \subfloat[Page Views] {
        \includegraphics[width=.31\textwidth]{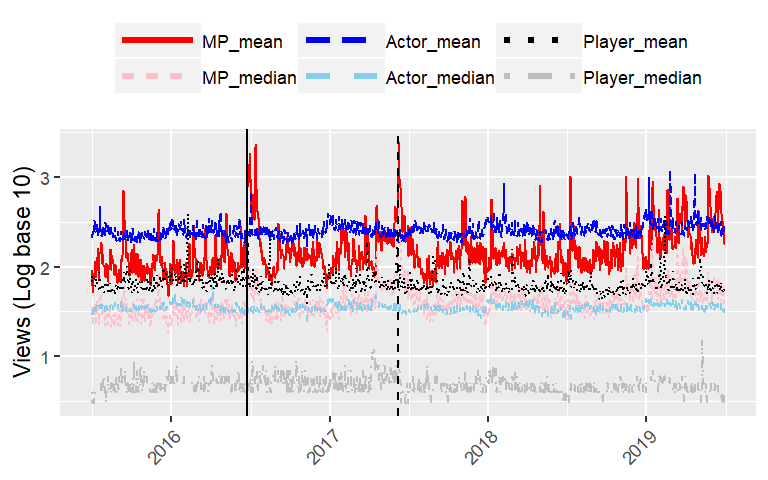}
        \label{fig:views}
    }
    \subfloat[Constituency Edits] {
        \includegraphics[width=.3\textwidth]{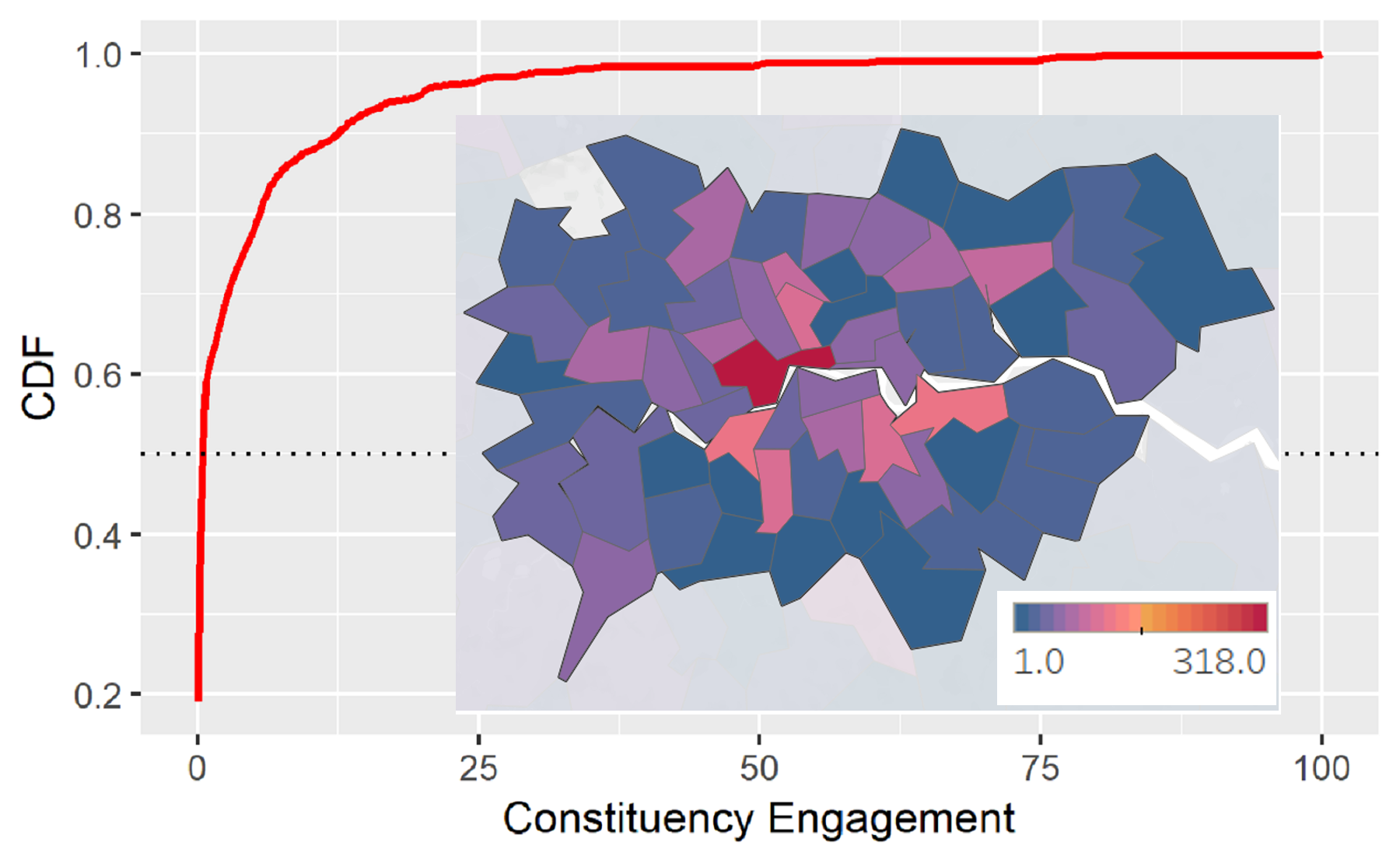}
        \label{fig:map}
    }
    \caption{Spatio-temporal patterns. 
    (a) CDF (Cumulative Distribution Function) of number of views, edits and editors per MP.
    (b) Page views of MPs pages and baselines (footballers and  actors in the UK). 
    (c) CDF of constituency engagement. Inset: All edits from constituencies of Greater London.}
\end{figure*}

Following are details of our dataset of Wikipedia articles for the 650 MPs elected in the 2017 general election. 
We summarise the distribution of the dataset in Figure~\ref{fig:stats}.

\textbf{Page Views} 
To understand readers' engagement, we collect the daily page views data on all articles, using Wikimedia's~\footnote{The Wikimedia foundation host projects and websites such as Wikipedia.} page view API.\footnote{\url{https://tools.wmflabs.org/pageviews/}, Accessed 26 Feb 2020} 
We use the earliest possible day that can be set (i.e. 01 July 2015) in the Wiki API query, and obtain daily page views per MP page until 30 June 2019.  In total, we observe over 160M views for 650 pages during this period. 

\textbf{Page Edits} 
We crawl the history of page edits for all 650 MPs from 01 June 2002 to 28 Aug 2019. We store in total 231k edits. These edits are made by 43k unique editors.
Across all edits we see that 55k edits are made using public IP addresses, which is shown as the username for anonymous editors. 

\textbf{Page Content}
To understand the content of the pages, we also collect page text as HTML dumps as on 18 July 2019. 
From these dumps, we extract the paragraph text, section titles, the citations list and other metadata for each page. 

\textbf{MPs Information} 
We obtain additional profile information of MPs using Wikidata.org, a free Wikimedia foundation knowledge base.
Wikidata provides information about MPs' gender, party, year of page creation and position held in the Parliament. 
We identify the role of each MP in politics, which is Wikidata's \emph{Position held (P39)} property.
For example, for the current prime minister \emph{Boris Johnson} his positions held include: \emph{Mayor of London, Secretary of State, Member of Parliament, Prime Minister and Leader of the Conservative Party}.

\textbf{Additional Data}
We obtain additional baseline data from Wikipedia and Twitter. We collect MPs' interactions on Twitter, such as number and popularity of mentions across 2 months from~\cite{agarwal2019tweeting}. For Wikipedia, we crawl page views for \emph{Sportspeople} (footballers) and \emph{actors}. These categories are popular biographies of living people in the UK\footnote{\url{www.wikidata.org/wiki/Wikidata:Living_people/uk},  Accessed 15 Feb 2020}. We randomly sample 1000 pages from a Wikidata page list of these two categories.

\section{Spatio-temporal patterns of Pages}\label{sec:dynamics}

We begin by investigating when the MPs' Wiki pages are created and edited and
what events or actions have impact on page views. In summary, we find that most MP pages are created soon after they are first elected; edits happen after significant changes such as elections or scandals; and most views happen just before or after important events such as referenda, elections and scandals.

\subsection{Page Views}
Using edit history for each article, we obtain the article creation date (i.e. the day when the first edit was made).  The edit history data covers four UK general elections (2005, 2010, 2015 and 2017). We see from that for the majority of pages, the creation date is close to a general election (60\% of page creation falls within these four years). In addition, 25\% of the articles were created between 2002 and 2004 (both included), coinciding with the birth and subsequent rapid growth of Wikipedia.
We focus on the dynamics of viewing behaviour. MP pages obtain a large number of views, with an average of 247k views per MP during the period we consider. 
We show the average view count of each day (mean) and median in Figure~\ref{fig:views} (Note: Y axis is log scale). The mean views count is high on days of two major events: the \emph{UK EU membership referendum} (also known as the `Brexit' referendum) of 24 June 2016 and the \emph{UK general election} of 09 June 2017. 

We note from Figure~\ref{fig:views} that near the Brexit referendum there are multiple viewership peaks. This relates to a major reshuffle of government ministers, which took place after Theresa May became the UK's new prime minister. This is similar to the patterns exposed in ~\cite{gobel2018political} for German MPs, where highest views generally follow ministers' resignations and new appointments.
During the \emph{UK general election} we see that the number of views rises from the day of the election announcement~\cite{election2017}.

Apart from events where a majority of MP pages get attention, there are events which are specific to individual (non-popular) MPs when they are in the limelight. Examples include anticipated and unanticipated events, such as ministers' resignations, speeches, interviews etc.~\cite{agarwal2019tweeting}.


\subsection{Page Edits and their Spatial Distribution}
Another core dimension of engagement with Wikipedia is the editing process carried out by largely anonymous volunteers. 
Among all edits, 36\% happen during the last three election years captured in the dataset (2010, 2015 and 2017). These patterns are similar to the views pattern in Figure~\ref{fig:views} (Pearson's correlation: 62\%, $p<0.001$), and we omit the figure with daily edit counts due to limited space. 

Out of all the edits, around 55k (22\%) edits are by public IPs which are recorded in the page revision history for not logged-in (anonymous) users. 
To understand the spatial distribution of editors, we map each public IP to a possible physical location (postcode), 
using the service provided by \emph{db-ip.com}, a geo-location database.
We observe that 84\% of public IP edits are from the UK. The remainder are from countries such as the United States--1768 edits, Ireland--442 edits, Australia--364 edits, Canada--297 edits, etc.

We then map each postcode to a local constituency using the list provided in~\cite{agarwal2019tweeting}.  We plot edit location at constituency level in Figure~\ref{fig:map} (inset), and observe that most public IP edits come from the area of Greater London, and more specifically from the Westminister borough (318 edits) where the Parliament sits. This is indicative of MPs' staff managing edits, similar to the findings of previous studies, which could provide a source of bias in the articles~\cite{gobel2018political}. 
The highest number of public IP edits coming from an MP's local constituency is for MP \emph{Amber Rudd MP}, former home secretary, who has 47 out of 232 edits from her own constituency (Hastings and Rye). 

To further quantify the extent to which edits come from an MP's constituency, we calculate the \textit{Constituency Engagement Factor (CE)} of each MP $m$ as the proportion of the edits to their articles which are localised to their constituency. If $m$ has $N_{m}$ edits of their page and $N_{m,c}$ edits from their constituency, we write: 
\begin{equation}\label{eq:localengagement}
    CE_{m,c} = \frac{N_{m,c}}{N_m}
\end{equation}
We plot the distribution of the metric in Equation~\ref{eq:localengagement} in Figure~\ref{fig:map}. We see that for 31\% of MPs there is at least one edit from their own constituency, but $CE$ is in general low, and only 6 MPs have more than half of their edits from their own constituency.



\section{Polarisation and Quality}\label{sec:polarization}

We continue our study by looking at information quality through the lenses of potential ideological and societal biases in Wikipedia articles about UK MPs. We ask and answer two questions: \textit{(i)} Do editors have an ideological bias, e.g., focusing on editing pages of MPs from a specific party? \textit{(ii)} What is the quality of citations used for MP pages? Do they have an ideological slant, and is their coverage sufficient? We find that there is a specialisation of editors, with some focusing mainly on Conservative party MPs, others on Labour MPs, and so on. 

\subsection{Editors' Ideological Preference}
To understand the extent to which editors tend to polarize around a specific ideology, we start by tracing, for each of the 42k editors, the party of pages which they mostly edit.
To this aim, we compute the number of edits to MPs from a given party, similar to~\cite{barbera2015birds,agarwal2019tweeting}, and associate each editor to the party to which they contribute the maximum edits. 
We do this for editors editing at least three different MPs, in order to exclude cases in which an editor is 
only interested in one MP or two MPs from different parties. This filtering step leaves us with 4.2k (7\%) of editors who are collectively responsible for 67\% of edits. 


\begin{figure}[thb]
    \centering
    \subfloat[Editors Network] {
     \includegraphics[width=0.4\columnwidth]{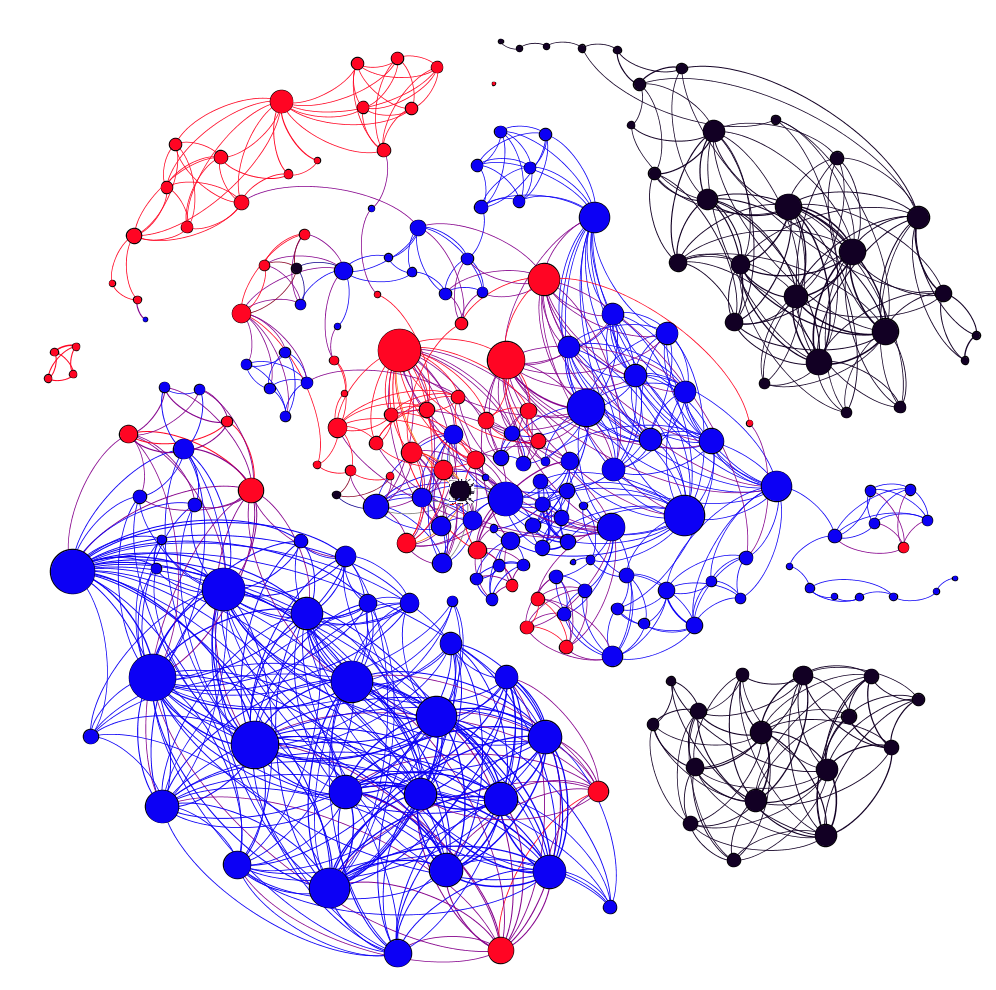}
    \label{fig:crossPartyNet}
    }
    \subfloat[Ideology scores] {
    \includegraphics[width=0.57\columnwidth]{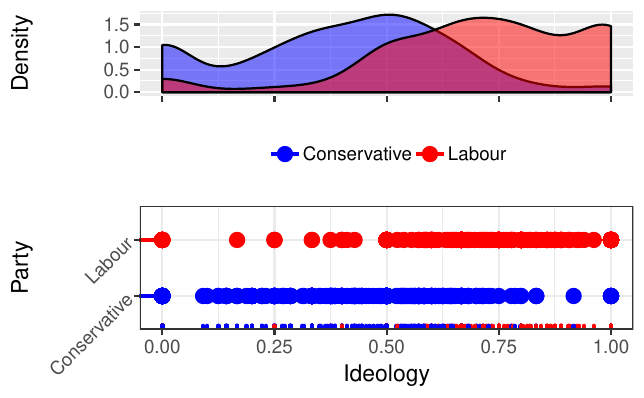}
    \label{fig:ideology}
    }
    \caption{Measure of Polarisation (Red: Labour, Blue: Conservative, Black: Others). 
    (a) Network graph based on editors as nodes, with edges connecting editors who have edited the same MPs' pages. (b) Ideology scores and density based on citations domains.}
\end{figure}

To understand communities forming this polarisation we perform network analysis -- 
 we define editors  as nodes and induce weighted edges between each pair of editors by computing the \emph{Jaccard Coefficient} or similarity of the sets of pages edited by them. Thus, if two editors
edit exactly the same set of MP pages, the weight will have its highest value of 1, and if there is no overlap, the value will be 0 (considered as no edge). We then use the Louvain method to identify communities of editors who have more connections within each community, but not many connections across communities. We find that this graph cleaves into 8 tightly knit communities, with a high modularity score of 0.229, indicative of polarisation or specialisation by party among editors. 

To visualise this better, we focus on the most active editors, and remove nodes with a degree of less than 5. To remove clutter, we do not show edges with weights less than 0.5.  Figure~\ref{fig:crossPartyNet} depicts this graph  by colouring the nodes (editors) based on their party (Blue for Conservative editors, red for Labour and black for others), and visually confirms the polarisation detected above by showing how the graph of editors divides along party lines,.  




\subsection{Citations Preference}
We next focus on the \emph{content} rather than the authors. We may expect that if authors exhibit polarisation, the sources they draw from to write the content, i.e., the MP pages, may also be polarised. To quantify this, we use URLs which are embedded in the \emph{References} section as citations. We find that there are 19k citations on MP pages. The average citation count is 29 per page and the median is 19. By checking the top citation URLs, we see that majority of them are news domains. This is consistent with the English Wikipedia articles study by~\cite{piccardi2020quantifying} which shows that top domains cited are 
Google.com, DOI.org, Nytimes.com. NIH.gov, BBC.co.uk, TheGuardian.com and so forth. Similar to~\cite{piccardi2020quantifying} we also extract the base domain from each citation URL and obtain 2212 unique domains. Additionally we find long URLs for 1.4k web archive \emph{(web.archive.org and archive.is)} short URLs. The top 5 domains which we get are BBC (15\%), theguardian (11\%), telegraph (6\%), parliament.uk (6\%), independent (5\%). The top 10 domains cover 52\% of the citations list. 

To check the use and representation of news domain sources on MP pages we compute an ideology score for each MP page. To do this we first check and label all possible news domains with their political leaning. We use scores from \emph{Mediabiasfactcheck}~\cite{mediaBias} and label the top 50 news domains with their political leaning (left, center or right). We also add ideology scores for news domains using sub-strings such as conservative (right-leaning) or labour (left-leaning) in the domain name. With this approach we find and label 67\% (13k) of the domains. 
Some examples of top domains by citations count are Left--Theguardian (2069), Independent (969), Newstatesman (262); Center--BBC (2766), Parliament.uk (1151), UKwhoswho (233); and Right--Telegraph (1207), thegazette (302), thetimes (251). 



After labeling, we compute an ideology score for each page, as the fraction of identified and \textit{Mediabiasfactcheck}-labelled news domains on that page that are left-leaning (i.e., ideology score = (number of left leaning news domains in citations)/(number of left+right leaning citations)). Note that a score of 0 indicates usage of only right-leaning sources and a score of 1 indicates only left-leaning sources; a score of 0.5 indicates a perfect balance.  Figure~\ref{fig:ideology} shows the ideology scores of Conservative and Labour pages as scatter and density plots. We see two peaks (median Labour: 0.7 and median Conservative: 0.4) which indicate a slight polarisation, with a slightly more polarised (i.e., farther away from 0.5) score for Labour MP pages. Also, 53 (17\%) Conservative and 57 (22\%) Labour pages have extreme polarised scores of 0 and 1 respectively. The KS statistics test (two sample) also confirms that there is a significant difference ($p<0.001$) with a distance value $D=0.54$ in the two parties' ideology scores. 

\subsection{Citation Quality}
To further understand the quality of Wikipedia articles, we perform an additional experiment on citations,  a key element for monitoring information quality in Wikipedia. 
Are MP articles well sourced? To answer this question, we compute a citation quality score for each page. 
To this end, we employ the ``Citation Need'' model defined in ~\cite{Redi:2019:CNT:3308558.3313618}. The model takes as input a sentence in a Wikipedia article and assigns a Citation Need score $s_{cn}$ in the range $[0,1]$. The higher the Citation Need score, the more likely it is that the sentence needs a citation.

\begin{figure}[th]
    \centering
     \includegraphics[width=0.9\columnwidth]{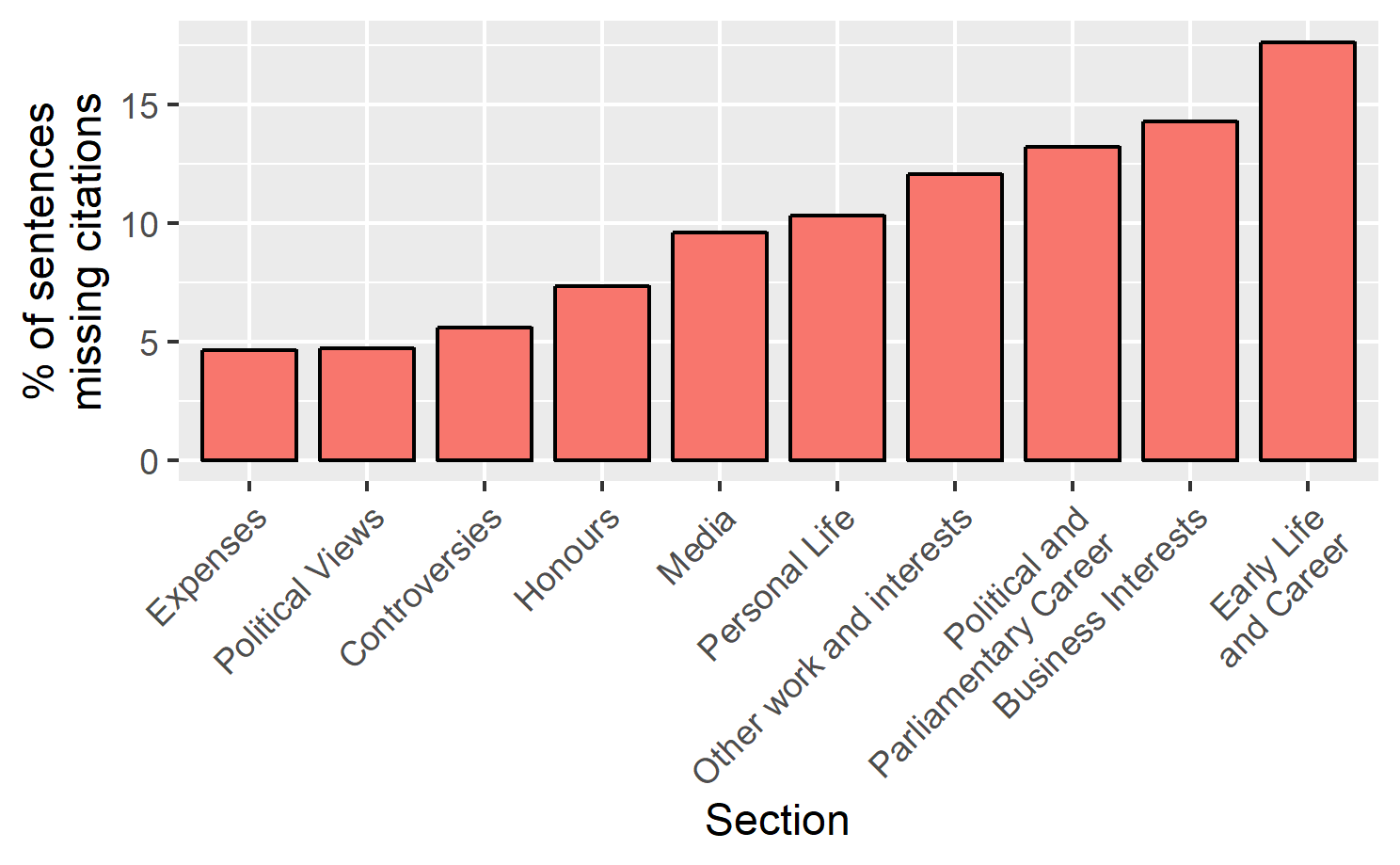}
 \caption{
Percentage of sentences which miss citation.}
\label{fig:citation}
\end{figure}

For each article we first parse all sentences and score them with the Citation Need model, after  filtering out statements in the  main section, which are less likely to need a citation
~\cite{Redi:2019:CNT:3308558.3313618}. 
Next, we aggregate the sentence-level scores and compute a Citation Quality $CQ$ as the proportion of sentences needing citations that already have a citation. To do so, we first identify the set ${C}$ of sentences needing citations in an article ($N$), namely those sentences whose $s_{cn}$ is greater than 0.5. Next, we count how many sentences in the set ${C}$ already have a citation in the original article text ($N_c$). Finally, we compute citation quality as the ratio between these two quantities:
\begin{equation}
CQ=\frac {N_c}{N},
\end{equation}

We check the distribution of citation quality values for each party. 
We see with the KS statistics test that for all party pairs there is no significant difference. However, we see that some pages for all parties have values of less than 0.5. We manually check these pages (50) and find that they are mostly in the Stub, Start and C (has significant problems) quality categories.

We see that the average citation quality for MP articles is pretty high: about 83\% of sentences needing citations actually have one. This is higher than the average citation quality in English Wikipedia, and comparable to articles about Medicine and Biology, which historically are constantly monitored for information quality \footnote{\url{meta.wikimedia.org/wiki/Research:Mapping_Citation_Quality}}.
We see that citation quality is uniform across MPs from different parties.
Among the articles with high citation quality score we find many with low quality Wikipedia article scores: for example, Scottish MPs David Linden and Martin Docherty, despite having just a few lines in their article (quality ``Start''), have a $CQ$ equal to 1 because of the richness of their citations. Conversely, we find that most articles with low citation quality are also low quality articles, with a few exceptions including the article about MP Robin Walker.

We report the percentage of statements where citation is missing in given sections in Figure~\ref{fig:citation}. For instance we see that sentences from \emph{Early Life and Career} miss citations the most often (18\%). It is likely that this section can be hard to find citations for as there may be fewer references in the digital media. We see that Expenses, Political Views, Controversies and Honours are the best cited sections with only 5\%--7\% of statements missing citations.

\section{Discussion}\label{sec:discussion}
In this paper, we discussed important dynamics of attention and content production in the context of Wikipedia articles focusing on UK Members of Parliament. We find evidence of specialization in contribution patterns of the editors of MP pages. 
With our analysis, we contribute to the broader field of online political communication studies, and shed light on behavior of contributors to the largest online encyclopedia.
We summarise our findings as follows:

\textbf{Spatio-Temporal Peaks of Engagement.} Similar to previous work, we find that MP page creations, views and edits are strongly aligned with media coverage and election periods. Furthermore, we find that only a small fraction of edits come from within the constituency of the MP, whereas the majority of anonymous edits come from Central London where the political centre of the UK lies. Collectively, these findings suggest that attention peaks are localized both in time and space, thus introducing potential vulnerabilities to the integrity of the content. Researchers working on monitoring and detecting coordinated disinformation attacks in political communications might benefit from these findings and further investigate how these peaks of attention might affect temporarily the quality of content on Wikipedia MP pages. 

\textbf{Polarisation Dynamics.}  We observe signals of partisanship among edits and domains of citations on papers. Many editors  contribute to the pages of MPs of one party, and we also see communities of editors who collectively focus on each party. We also see a different distribution of citation ideology scores between pages of MPs from the two main UK parties, with Labour MP pages drawing on sources with a slightly higher degree of ideological bias. To further investigate these findings, we contacted experienced editors from the sub-community of Wikipedia curating pages of UK MPs. One possible explanation for this polarisation of sources is that, before election, Labour MPs tend to be less well known than Conservative MPs, and thus only left-wing press tend to cover them when they are nominated. This evidence of polarisation therefore might not necessarily imply political bias, but further research in this direction should  investigate the neutrality of the content coverage in those articles.

\textbf{High Source Quality.} In terms of verifiability and information quality, we find that MP articles are generally well sourced, with an average of only 10\% of sentences detected as missing citations. The citation quality is therefore very high overall, similar to Wikipedia medical articles, with sections on early life and career having the highest proportion of statements which lack citations. This suggests that, despite signals of polarisation among contributors, the information in Wikipedia articles is backed by a sufficient amount of sources. Our study finds that the UK Parliament's website is the source of (6\%) of citations on these pages. Official sources like this can be used to improve quality and trustworthiness, although they will not provide effective sources for aspects such as early life and career.


\textbf{Limitations and Future Work} In this paper, we do not investigate the  dynamics of page evolution over time. We use the final stable page state as of 28th Aug 2019 to compute metrics and features for the content and time patterns. Future work should focus on mining temporal patterns of engagement, including changes resulting from the recent 2019 general election and presence of pages in multiple languages~\cite{hale2014multilinguals,agarwal2020characterising}. 
Also, while we trace quality patterns across different topics, we do not look at the value of individual contributions, which could be used to help identify spam and vandalism and thereby measure online hate or disinformation campaigns. 


\section{Acknowledgements}
We also acknowledge  support via  EPSRC Grant Ref: EP/T001569/1 particularly the theme for ``Detecting and Understanding Harmful Content Online: A Metatool Approach'', as well as a Professor Sir Richard Trainor Scholarship 2017 at King's College London.

\balance{}

\bibliographystyle{unsrt}
\bibliography{references}

\end{document}